\documentclass[twocolumn,prb, showpacs]{revtex4}
\usepackage{psfrag}
\usepackage{graphicx}
\usepackage{amsmath}
\usepackage{amssymb}
\usepackage{eufrak}
\usepackage[ansinew]{inputenc}

\begin{document}

\title{A Higgs mass at 125 GeV calculated from neutron to proton decay in a u(3) Lie group Hamiltonian framework}

\author{Ole L. Trinhammer}
\affiliation{Department of Physics, Technical University of Denmark, \\
Fysikvej, Building 307, DK-2800 Kongens Lyngby, Denmark, EU.\\} 

\begin{abstract}
  We investigate the neutron to proton decay via a Higgs mechanism in the framework of a reinterpreted Kogut-Susskind Hamiltonian on the Lie group u(3). We calculate expressions for a scalar Higgs mass, an electroweak energy scale, and vector gauge boson masses which all compare well with observed or derived values. Our sole ad hoc inputs to the calculations are the classical electron radius and the weak mixing angle. Our result for the Higgs mass relative to the electron mass involves only mathematical constants and the fine structure constant. It yields 125.1 GeV for a fine structure constant taken as a geometric mean between it's sliding scale values at respectively the electron mass and the W vector boson mass which are both involved in the neutron decay. In passing we compare with the neutral flavour baryon spectrum and mention an approximate calculation of the relative neutron to proton mass ratio of 0.13847 percent which is promisingly close to the observed value of 0.137842 percent. We finally mention the Fermi coupling constant as a derived quantity. 
\end{abstract}


\maketitle

\section*{1 Introduction} 

We have previously presented a hamiltonian structure for baryonic states with configuration variable $u=e^{i\chi}$ in the Lie group $u(3)$ [\cite{TrinhammerEPL102}]. Our Hamiltonian reads
\begin{equation} \label{eq:hamiltonian}
   H=\frac{\hbar c}{a}\left[-\frac{1}{2}\Delta+\frac{1}{2}\rm{Tr}\ \chi^2\right]
\end{equation}
with an energy scale $\Lambda\equiv\ \hbar c/a\approx 215\ \rm{MeV}$ corresponding to a length scale $a$ which we related to the classical electron radius $r_e=e^2/(4\pi\epsilon_0 m_e c^2)$ [\cite{Heisenberg}, \cite{LandauLifshitz}, \cite{RPP2012p107}] by a projection $\pi a=r_e$ [\cite{TrinhammerEPL102}]. It is not a new idea to use the classical electron radius as a scale for strong interaction phenomena [\cite{Heisenberg}]. In the present work we shall study our projection in the framework of a Higgs mechanism adopted from the standard model for the electroweak interactions [\cite{RPP2012pp136}, \cite{WeinbergQToFIIpp305}, \cite{McMahonQFT}, \cite{AitchisonHey}, \cite{AitchisonHey4thVol2p380}]. The goal is to reduce the number of ad hoc parameters in the description of baryon phenomena. Here we restrict our inputs to the above mentioned scale and the weak mixing angle $\theta_W$. In that way we can express - in terms of the fine structure constant $\alpha=e^2/(4\pi \epsilon_0\hbar c)$ [\cite{RPP2012p107}, \cite{Sakurai}] and the weak mixing angle - all masses coming out of the model relative to the value of the electron mass. Our results are for the electroweak energy scale $v$ [\cite{AitchisonHey4thVol2p380}], the Higgs scalar mass $m_H$, and the electroweak gauge vector boson masses $m_W$, $m_Z$
\begin{gather}
  v=2\sqrt{2}(\frac{\pi}{\alpha})\Lambda=2\sqrt{2}(\frac{\pi}{\alpha})^2m_ec^2=250\ \rm{GeV} \nonumber \\
    m_Hc^2=\sqrt{2}\frac{\pi}{\alpha}\Lambda=\sqrt{2}(\frac{\pi}{\alpha})^2m_ec^2=125.1\ \rm{GeV} \nonumber \\
  m_Wc^2=\sqrt{\frac{4\pi\alpha}{\sin^2\theta_W}}\sqrt{2}(\frac{\pi}{\alpha})^2m_ec^2=80.1\ \rm{GeV} \nonumber \\
  m_Zc^2=\sqrt{\frac{4\pi\alpha}{\sin^2\theta_W\cos^2\theta_W}}\sqrt{2}(\frac{\pi}{\alpha})^2m_ec^2= 91.4\ \rm{GeV}. \label{eq:improvednumerics}
\end{gather}
The numerical results are for the fine structure constant taken as a geometric mean $\hat{\alpha}$ from sliding scale values [\cite{RPP2012pp136}, \cite{WeinbergQToFIIp158p126}, \cite{LandauLifshitzQED}] between electronic and vector boson energies,
\begin{equation}	\label{eq:alphamean}
  \hat{\alpha}^{-1}=1/\sqrt{\alpha(m_e)\alpha(m_W)}=132.42.
\end{equation}
For the weak mixing angle we have used $\sin^2\hat{\theta}(m_Z)=0.23116$  at $m_Z$ [\cite{RPP2012p107}].
Just to see the sensitivity from $\alpha(m_W)$ to $\alpha(m_Z)$ in (\ref{eq:alphamean}) we mention that using $\alpha(m_Z)^{-1}=127.944$ [\cite{RPP2012p137}] in (\ref{eq:alphamean}) we get $\hat{\alpha}^{-1}=132.41$ whereas there are no changes in (\ref{eq:improvednumerics}) at the level of significant digits given there. The two latter results in (\ref{eq:improvednumerics}) are to be compared with the experimental values $m_Wc^2=80.385(15)\ \rm{GeV}$ and $m_Zc^2=91.1876(21)\ \rm{GeV}$ cited in [\cite{RPP2012p107}]. Below we briefly describe our baryon model and introduce the Higgs mechanism leading to the above results.

\section*{2 Expansion on Bloch wave Slater determinants}

\begin{figure} 
\begin{center}
\includegraphics[width=0.45\textwidth]{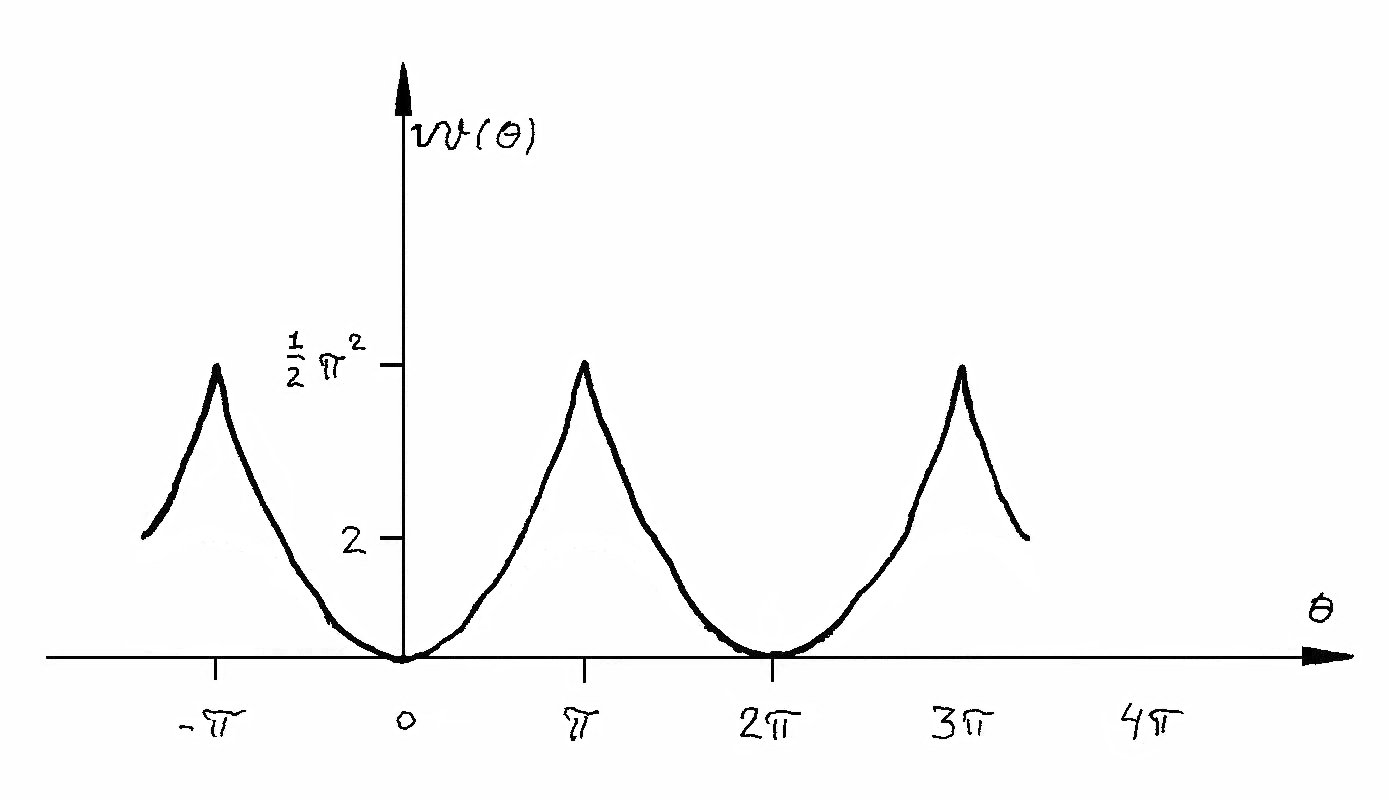}
\caption{Periodic parametric potential (\ref{eq:periodicpotential}) as a function of eigenangles of our u(3) configuration variable.}
\label{fig:periodicpotential}
\end{center}
\end{figure}

\begin{figure} 
\begin{center}
\includegraphics[width=0.45\textwidth]{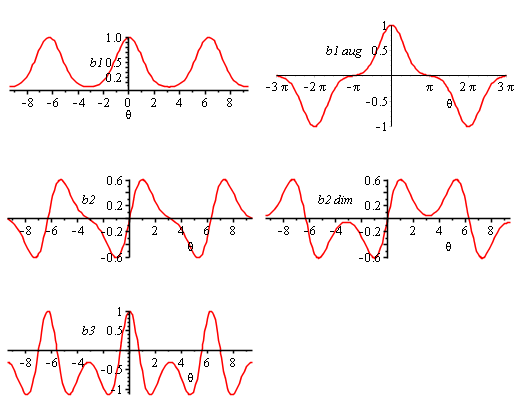}
\caption{Parametric eigenfunctions from (\ref{eq:onedimSchoedinger}). The period doubling (right) in the diminished state for level two is paired with an augmented period doubled state for level one (above).}
\label{fig:parametric}
\end{center}
\end{figure}

\begin{figure}
\begin{center}
\includegraphics[width=0.35\textwidth]{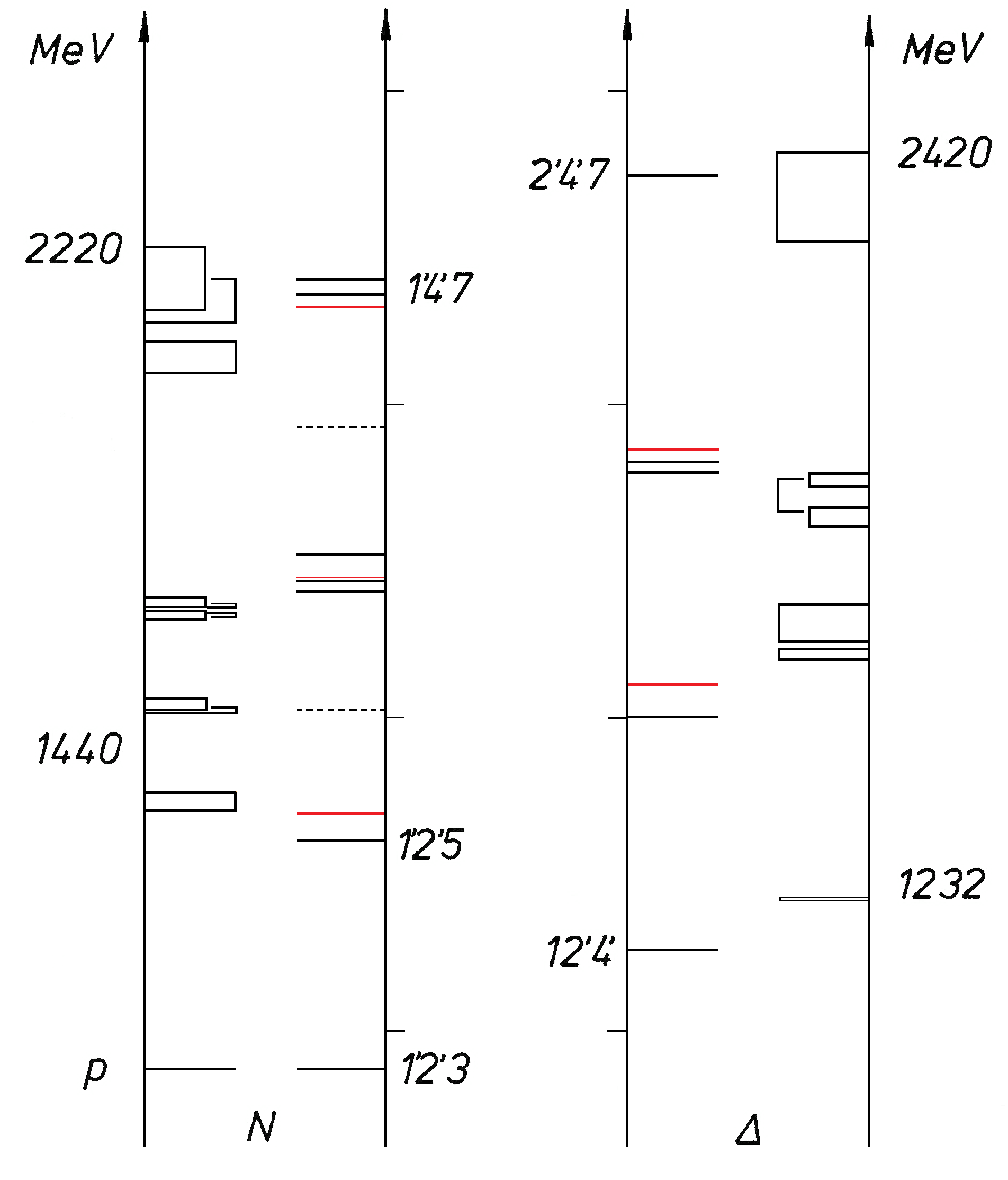}
\caption{All observed certain neutral flavour baryons (boxes) compared with approximate predictions (black, red and dashed lines) from eq. (\ref{eq:SeperableSchroedinger}). The dashed lines represent neutral flavour singlets, particular for the present model. The red lines mark states with augmented contribution in level 3. The boxes indicate the experimental range of pole positions [\cite{RPP2012}], not the resonance widths which are much larger. We have made no estimate of mass shifts due to strong coupling to decay channels [\cite{Hoehler}]. Digits at selected predictions are parametric labels $p, q, r$ based on table I. Note the fine agreement in the grouping and the number of resonances in both sectors.}
\label{fig:NDeltaSpectrum}
\end{center}
\end{figure}

Our configuration space is the Lie group $u(3)$ where the elements $u$ have eigenvalues $e^{i\theta_j}$ and the three independent eigenangles $\theta_j$, $j=1,2,3$ are real. Our reinterpreted Kogut-Susskind [\cite{KogutSusskind}, \cite{Trinhammer1983}] Hamiltonian (\ref{eq:hamiltonian}) operates on states $\Psi(u)$ and we expect the implied Schr\"{o}dinger equation
\begin{equation} \label{eq:schroedinger}
   \frac{\hbar c}{a}\left[-\frac{1}{2}\Delta+\frac{1}{2}\rm{Tr}\ \chi^2\right]\Psi(u)=E\Psi(u)
\end{equation}
to describe the baryon spectrum e.g. as shown in fig. \ref{fig:NDeltaSpectrum} below for neutral flavour as constructions from table I. Equation (\ref{eq:schroedinger}) can be parametrized by using a polar decomposition of the Laplacian on $u(3)$ [\cite{TrinhammerEPL102}, \cite{TrinhammerOlafsson}]. The ground state for unbroken symmetry is identified with the neutron [\cite{TrinhammerEPL102}] and the dimensionless eigenvalue ${\rm{E_n}}\equiv E_n/\Lambda$ can be found by a Rayleigh-Ritz method to yield $\rm{E_n}=4.3820$ for an expansion on 3078 base functions. In the present work we shall focus on the shortest geodetic trace potential $\frac{1}{2}d^2(e,u)=\frac{1}{2}\rm{Tr}\ \chi^2$ with $e=I$ the {\it origo} of $u(3)$. In parameter space it folds out as
\begin{equation}
  \frac{1}{2} {\rm{Tr}}\ \chi^2\equiv W =w(\theta_1)+w(\theta_2)+w(\theta_3),
\end{equation}
in a sum of periodic parametric potentials, see fig. \ref{fig:periodicpotential}
\begin{equation} \label{eq:periodicpotential}
  w(\theta)     = \frac{1}{2}(\theta - n\cdot 2\pi)^2, \quad \theta \in [(2n-1)\pi,(2n+1)\pi], n\in \mathbb{Z}.
\end{equation}
The geodetic distance potential can be used as an interaction term in a model for two baryons with configuration variables $u$ and $u'$ for which $d(u,u')=d(e,u'u^\dagger)$. Thus we conjecture the deuteron to be the ground state of
\begin{equation}
   \frac{\hbar c}{a}\left[-\frac{1}{2}\Delta_u-\frac{1}{2}\Delta_{u'}+\frac{1}{2}d^2(u,u')\right]\Psi(u,u')=E\Psi(u,u')
\end{equation}
When one imagines a projection of the term $u'u^\dagger$ it does have the "smell" of a quark-antiquark structure characteristic of mesons. However we do not want to pursue this idea here. Returning to single baryons the wave function $\Psi$ in (\ref{eq:schroedinger}) is factorized into a toroidal part $\tau$ and an off torus part $\Upsilon$
\begin{equation}
  \Psi(u)=\tau(\theta_1,\theta_2,\theta_3)\Upsilon(\alpha_4,\alpha_5,\alpha_6,\alpha_7,\alpha_8,\alpha_9).
\end{equation}
In that way (\ref{eq:schroedinger}) can be solved for specific choices of spin and flavour inflicted by the six off torus generators contained in the Laplacian. After integration over the six off-toroidal degrees of freedom $\alpha_4,\alpha_5,\alpha_6,\alpha_7,\alpha_8,\alpha_9$ one ends up with a Schr\"{o}dinger equation
\begin{equation}    \label{eq:toroidalSchroedinger}
 [ -\frac{1}{2}\sum^3_{j=1} \frac{\partial^2}{\partial \theta_j^2}+V]R(\theta_1,\theta_2,\theta_3) = \mbox{E}R(\theta_1,\theta_2,\theta_3).
\end{equation}
Here $R=J\tau$ is the toroidal wavefunction scaled by the 'Jacobian', the van de Monde determinant [\cite{Weyl}]
\begin{equation}    
      J = \prod^3_{i <  j } 2 \sin\left(\frac{1}{2} (\theta_i- \theta_j)\right)
\end{equation}
from the polar decomposition of the Laplacian and
\begin{gather}    
  V = -1 + \frac{1}{2}\cdot\frac{4}{3}\sum^3_{ i <  j} \frac{1}{8 \sin^2 \frac{1}{2}(\theta_i -\theta_j)}
\nonumber \\\hspace{1cm} +  w(\theta_1) + w(\theta_2)+ w(\theta_3). \hspace{2.5cm}
\end{gather}
contains in the second term contributions from off-toroidal degrees of freedom that carry spin and flavour in the specific choice here of spin, hypercharge and isospin $s=1/2, y=1, i=1/2$. The measure scaled wavefunction $R$ can be expanded on solutions $b$ to the separable problem
\begin{equation}    \label{eq:SeperableSchroedinger}
 [ -\frac{1}{2}\sum^3_{j=1} \frac{\partial^2}{\partial \theta_j^2}+W]b(\theta_1,\theta_2,\theta_3) = \mbox{E}b(\theta_1,\theta_2,\theta_3).
\end{equation}
Such solutions can be constructed as Slater determinants [\cite{Slater}]
\begin{equation}
  b_{pqr}=\epsilon_{ijk}b_p(\theta_i)b_q(\theta_j)b_r(\theta_k),
\end{equation}
where $p,q,r$ are natural number labels for orthogonal solutions to the one-dimensional Sch\"{o}dinger equation
\begin{equation} \label{eq:onedimSchoedinger}
  [ -\frac{1}{2}\frac{\partial^2}{\partial \theta^2}+w(\theta)]b_p(\theta)=e_pb_p(\theta)
\end{equation}
with periodic parametric potential. Figure \ref{fig:parametric} shows solutions for the first three eigenvalues $e_1, e_2, e_3$.
\begin{figure}
\begin{center}
\includegraphics[width=0.45\textwidth]{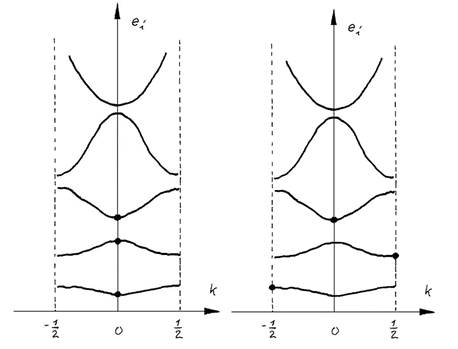}
\caption{Reduced zone scheme [\cite{AshcroftMermin}] for parametric eigenvalues. The black dots represent the values for the unstable neutron state (left) and the proton state (right). For clarity the variation of the eigenvalues with Bloch wave number $\kappa$ is grossly exaggerated for the lowest states.}
\label{fig:reducedzone}
\end{center}
\end{figure}

The structures of (\ref{eq:SeperableSchroedinger}) and (\ref{eq:toroidalSchroedinger}) with periodic potentials either $V$ or $W$ imply the introduction of Bloch wave expansion factors
\begin{equation} \label{eq:gp}
  g_p(\theta)=e^{i\kappa\theta}u_p(\theta),
\end{equation}
where $\kappa$ introduces the Bloch degree of freedom. We shall argue that the Bloch degrees of freedom are opened by a Higgs mechanism that will allow a diminishing of the ground state eigenvalue via the creation of the $\nu_e, e_L$ doublet and it's coupling to a Higgs field. For instance the ground state eigenvalue $\rm{E_n}=e_1+e_2+e_3=4.47...$ of (\ref{eq:SeperableSchroedinger}) is lowered to a value $\rm{E_p}=e'_1+e'_2+e_3=4.46...$ for real symmetry broken states of parametric eigenvalues $e'_1$ and $e'_2$ with $4\pi$ periodicity analogous to $\kappa_1,\kappa_2=\pm\frac{1}{2},\pm\frac{1}{2}$ for the Bloch phase containing $g$s as opposed to the $2\pi$ periodicity of the $b_p$s and $u_p$s, see fig. \ref{fig:reducedzone}. The period doublings are allowed since they leave the square of the wave function $\Psi^2$ singlevalued on $u(3)$. Table \ref{tab:paraeig} shows results for the parametric eigenvalues. Different combinations of three different levels give a good reproduction of the observed spectrum of all the certain (four star) neutral flavour baryon resonances, i.e. the N and $\Delta$ spectrum with {\it no missing resonance problem} [\cite{RPP2012p205}], see fig. \ref{fig:NDeltaSpectrum}. 

Summing up the three lowest levels we get an approximate estimate of the relative neutron to proton mass shift
\begin{equation}
  \frac{m_n-m_p}{m_p}\approx\frac{e_1+e_2+e_3-(e'_1+e'_2+e_3)}{e'_1+e'_2+e_3}=0.13847{\%}.
\end{equation}
This is to be compared with the value $0.137842{\%}$ calculated from the observed neutron and proton masses which are known experimentally with eight significant digits [\cite{RPP2012p79}]. The exact value for $\rm{E_n}$ from (\ref{eq:toroidalSchroedinger}) is $4.38...$ which is a few percent lower than the approximate value $\rm{E_n}=4.47...$ mentioned above. A suitable base on which to expand an exact calculation for $\rm{E_p}$ has not been found.
\begin{table}
\begin{center}
\caption{Eigenvalues of the parametric group space chopped harmonic oscillator Schr\"{o}dinger equation (\ref{eq:onedimSchoedinger}) calculated with 1500 collocation points. Note that the lowest eigenvalues as expected are close to those of the ordinary harmonic oscillator. Moving up to higher levels the eigenvalues differ more and more from those of the harmonic oscillator as indicated in fig. \ref{fig:reducedzone}. \vspace{3mm}}
\label{tab:paraeig}
\begin{tabular}{ c c c c } \hline\hline\\
$p$&$e_p$&$e'_p$&$e'_p$ \\
Level&Eigenvalue&Diminished&Augmented \\ \hline \\
1&0.499804708&&0.5001727904 \\
2&1.502988968&1.496433950& \\ 
3&2.471378779&&2.522629649 \\
4&3.600509000&3.377236032& \\
5&4.218515963&&4.803947527 \\
6&6.197629004&5.160535373& \\
7&6.383117406&&7.820486992 \\
8&9.688466291&7.922699154& \\ 
9&9.751335596&&11.80644676 \\
10&14.1755275&11.84897047& \\
11&14.2063708&&16.79575229 \\ \hline\hline
\end{tabular}
\end{center}
\end{table}

\section*{3 An exemplar Higgs mechanism for the neutron decay}
The eigenangles $\theta$ are local parametrizations of the $u(3)$ torus. In our model they are dynamical variables and as such can be identified as fields when projected to space, c.f. the role of $\pi$ fields in Skyrmion models  [\cite{DiakonovPetrovPolyakov}, \cite{DiakonovPetrovPobylitsaPolyakovWeiss}]. We tie up the strong interaction configuration and the electroweak sector by an {\it electroweak trailing} Ansatz
\begin{equation} \label{eq:HiggsAnsatz}
  \Lambda\theta=\alpha\phi(x)
\end{equation}
where $\phi$ is a Higgs field implied by the above mentioned period doubling degree of freedom. A vague analogy lies in the description of the decay of an excited state in an atom where the electron position coordinates act as dynamical variables. Here a spontaneous transition takes place in the electronic level structure through a coupling to the photon field, i.e. a photon is created at a random point and emitted in a random direction to exterior space. In our case the interior baryon configuration space is the analogue of the atom as an entire structure, now with non-spatial configuration variables that project to exterior space as field variables.

In a one-dimensional analogue of standard procedures in electroweak theory [\cite{{McMahonQFT}}] we allow for a local phase variation $e^{-iq\beta}$ in $\theta$ with $q$ soon to be identified with the electric charge coupling constant $e$. Thus we consider the local transformation
\begin{equation} \label{eq:localphase}
   \theta(x)\rightarrow \theta'(x)=e^{-iq\beta(x)}\theta(x).
\end{equation}
Following McMahon [\cite{{McMahonQFT}}] we have the standard Lagrangian 
\begin{equation} \label{eq:Lagrangian1ed}
   L=D_\mu\theta^\dagger D^\mu\theta-w(\theta)-\frac{1}{4}F_{\mu\nu}F^{\mu\nu}
\end{equation}
for $\theta$ and the local transformation by $\beta$ related to a gauge field $B_\mu$ 
\begin{equation}
   B_\mu\rightarrow B'_\mu = B_\mu+ \partial_\mu\beta.
\end{equation}
The gauge field renders the Lagrangian invariant under the transformation in (\ref{eq:localphase}) by the action of the generalized derivative
\begin{equation} \label{eq:covariantderivative}
   D_\mu\equiv \partial_\mu+iqB_\mu.
\end{equation}
We presume the neutron decay to be mediated through energetically favourable period doublings allowed in the periodic potential (6) for parametric eigenfunctions, see fig. \ref{fig:reducedzone}. These period doublings correspond to sudden jumps of $\theta$  from one trough to a neighbouring one. We therefore consider the restriction of $w(\theta)$  to a specific section like
\begin{equation} \label{eq:vshifted}
   w(\theta)=\frac{1}{2}(\theta-2\pi)^2
\end{equation}
neighbouring to the generic section at $n=0$. The section (\ref{eq:vshifted}) has a minimum at $2\pi$ and we investigate perturbations around this using (\ref{eq:HiggsAnsatz})
\begin{equation}  \label{eq:phishited}
  \phi \rightarrow\phi'=(\frac{2\pi}{\alpha}+\frac{2\pi}{\alpha}\frac{\rm{h}(x)}{\sqrt{2}})\Lambda=\frac{1}{\sqrt{2}}(v+h(x)).
\end{equation}
We have absorbed the scaled energy dimension $\overline{\Lambda}\equiv\Lambda 2\pi/\alpha$ into dimensionful entities $v\equiv\overline{\Lambda}\sqrt{2}\equiv\overline{\Lambda}\rm{v}$ and $h\equiv\overline{\Lambda}\rm{h}$ [\cite{WeinbergVolIp302}] and anticipated the factor $1/2$ in the mass term in the Lagrangian below by introducing the factor $1/\sqrt{2}$ in the expression above. The Lagrangian for the pertubing field $h(x)$ and the related gauge field $B'_\mu$ (now also scaled by $\overline{\Lambda}$) is found by inserting in (\ref{eq:Lagrangian1ed}) the eqs. (\ref{eq:covariantderivative}) and (\ref{eq:vshifted}) with $\phi$ substituted for $\theta$ according to (\ref{eq:phishited}) to give
\begin{widetext}
\begin{equation}
   L=\frac{1}{2}\partial_\mu h \ \partial^\mu h-\frac{1}{2}\frac{h^2}{2}+\frac{1}{2}(q{\rm{v}})^2B'_\mu B'^\mu+\frac{1}{2}q^2 h^2 B'_\mu B'^\mu+2q \frac{h}{\sqrt{2}}B'_\mu B'^\mu-\frac{1}{4}F_{\mu\nu}F^{\mu\nu}. 
\end{equation}
\end{widetext}
From the coefficients of the quadratic terms $h^2$ and $B'_\mu B'^\mu$ with $q=e=\sqrt{4\pi\alpha}$ in (\ref{eq:localphase}) and with dimensionless ${\rm{v}}=\sqrt{2}$ we read off the respective dimensionless masses $\rm{m_H}$ and $\rm{m_{B'}}$ determined by
\begin{equation} \label{eq:mHmBdimless}
   \frac{1}{2}\rm{m_H}^2=\frac{1}{2}\cdot\frac{1}{2} \ and  \ \frac{1}{2}\rm{m_{B'}}^2=\frac{1}{2}(\sqrt{4\pi\alpha}\sqrt{2})^2.
\end{equation}
Our length scale $a$ in the Hamiltonian originates in the classical electron radius mentioned in the introduction and thus our energy scale $\Lambda$ can be conveniently expressed in units of the electron mass $m_e$ by 
\begin{equation} \label{eq:me}
   \Lambda=\frac{\pi}{\alpha}m_ec^2.
\end{equation}
From the Higgs mechanism above we therefore get for the scalar field mass $m_H$
\begin{equation} \label{eq:mH}
   m_Hc^2={\rm{m_H}}\overline{\Lambda}=\frac{2\pi}{\alpha}\frac{1}{\sqrt{2}}\Lambda=\sqrt{2}(\frac{\pi}{\alpha})^2m_ec^2,
\end{equation}
which yields $m_H=125.1$ GeV for a geometric mean fine structure constant $\hat{\alpha}=(132.42)^{-1}$ between electron and vector boson energies. The expression (\ref{eq:mH}) containing solely the electron mass and the fine structure constant and cited in (\ref{eq:improvednumerics}) is determined by the trailing in (\ref{eq:HiggsAnsatz}) and by the structure of the potential (\ref{eq:vshifted}) and therefore remains valid below. Similarly we would get $m_{B'}c^2=78\ \rm{GeV}$. However for the vector gauge field masses corresponding to $m_{B'}$ we need to consider the full electroweak $SU(2)_L \times U(1)$ treatment to give the results in (\ref{eq:improvednumerics}).

\section*{4 A full two component Higgs mechanism}
The symmetry breaks introduced by the Bloch phase factors in the parametric eigenstates $g_p$ in (\ref{eq:gp}) have to come in pairs of half odd-integer valued Bloch wave numbers $(\kappa_1, \kappa_2)$ in order to "kill" the singularity in the centrifugal potential
\begin{equation}
  C=\frac{1}{2}\cdot\frac{4}{3}\sum^3_{ i <  j} \frac{1}{8 \sin^2 \frac{1}{2}(\theta_i -\theta_j)}.
\end{equation}
Generalizing our Ansatz we take the paired period doublings corresponding to the shift in fig. \ref{fig:reducedzone} from $(\kappa_1, \kappa_2)=(0,0)$ to $(\kappa_1, \kappa_2)=(\pm\frac{1}{2},\pm\frac{1}{2})$ to be mediated by a Higgs mechanism with a complex two-component doublet
\begin{equation}
  \phi= \begin{Bmatrix} \phi_1 \\ \phi_2 \end{Bmatrix}
\end{equation}
to "absorb" phase changes and a two-component electronic lepton 
\begin{equation}
  l_{eL}=\begin{Bmatrix} \nu_e \\ e \end{Bmatrix}_L
\end{equation}
to "take care" of  the remaining degrees of freedom. In common $\phi$ and $l_{eL}$ are topological moulds with the necessary degrees of freedom to "absorb" the kinematics of the liberated energy and to "mirror" the topological changes in the nucleon state following from the parametric period doublings in the neutron to proton transformation indicated in fig. \ref{fig:reducedzone} and (\ref{eq:gp}). Following Cornwell, Aitchison/Hey and Weinberg [\cite{Cornwell},  \cite{AitchisonHey4thVol2p380}, \cite{WeinbergQToFIIpp308}] we perform a transformation such that
\begin{equation}
  \phi= \begin{Bmatrix} \phi^+ \\ \phi^0 \end{Bmatrix}
\end{equation}
has the individual real valued component vacuum expectation values $<\phi^+>=0$ and $<\phi^0>\equiv v/\sqrt{2}$. For our case (\ref{eq:vshifted}, \ref{eq:phishited}) we have
\begin{equation} \label{eq:v}
  v=2\sqrt{2}\frac{\pi}{\alpha}\Lambda
\end{equation}
which relates the electroweak scale to the scale of the strong interactions and which can now be inserted into the standard results from the electroweak theory [\cite{AitchisonHey4thVol2p381}, \cite{WeinbergQToFIIp307p309}]
\begin{equation} \label{eq:mWandmZ}
  m_Wc^2 =\frac{v|g|}{2}, \ \ m_Zc^2=\frac{v\sqrt{g^2+g'^2}}{2}
\end{equation}
where the $SU(2)$ coupling constant $g$ and the $U(1)$ coupling constant $g'$ are given from the electric charge coupling constant $e=\sqrt{4\pi\alpha}$ and the electroweak mixing angle $\theta_W$ by
\begin{equation} \label{eq:gg'}
  g=-e/\sin\theta_W,\ g'=-e/\cos\theta_W.
\end{equation}
When (\ref{eq:v}), (\ref{eq:me}) and (\ref{eq:gg'}) are used in (\ref{eq:mWandmZ}) the results in  (\ref{eq:improvednumerics}) follow readily with no ad hoc $\phi^4$-term in the Higgs mechanism. With Weinberg-Salam [\cite{McMahonQFTp222}] we could assume a Yukawa coupling Lagrangian term [\cite{WeinbergQToFIIpp308}, \cite{AitchisonHey4thVol2p403}] to yield a direct determination of the Yukawa coupling constant $G_e=g_e/\sqrt{2}=m_ec^2/v$. It should further be noted that Weinberg's value for $G_e=2.07\cdot 10^{-6}$  follows from a determination of $v$ from the Fermi coupling constant $G_F$ whereas in our model $G_e=2.04\cdot 10^{-6}$ is determined directly by a sliding scale geometric mean $\hat{\alpha}^{-1}=132.42$ from (\ref{eq:alphamean}). In that way the Fermi coupling constant becomes a derived quantity
\begin{equation} \label{eq:GF}
  \frac{G_F}{(\hbar c)^3}=\frac{1}{\sqrt{2}}\frac{1}{v^2}=\frac{1}{8\sqrt{2}}(\frac{\alpha}{\pi})^4\frac{1}{(m_ec^2)^2}.
\end{equation}
It is not clear at which energy $\alpha$ should be taken in (\ref{eq:GF}). In (\ref{eq:improvednumerics}) we give results for a geometric mean between electronic and vector bosonic energies. In field theory terms our results are only to lowest order, thus the exact factor two between the electroweak scale $v$ and the Higgs mass $m_H$ is not supposed to hold in individual sliding scale treatments.  

\section*{5 Conclusion}
With just two ad hoc inputs - the classical electron radius which gives our scale and the weak mixing angle - we have investigated the neutron decay. Specifically we have investigated a Higgs mechanism for parametric period doublings of two coupled toroidal degrees of freedom in the configuration variables describing a shift from parametric eigenstates involved in a neutronic state to parametric eigenstates involved in a protonic state on the Lie group u(3). In that way we have related the strong and electroweak energy scales. The common scale of our Hamiltonian for baryonic phenomena expresses the scalar Higgs and vector gauge boson masses in units of the electron mass. The result for the Higgs mass is 125 GeV for a fine structure constant taken as a geometric mean between electronic to vector bosonic regimes. In passing we have related to a Fermi coupling constant prediction, given an approximate result without any ad hoc parameters for the relative neutron to proton mass ratio in promising agreement with the experimentally determined value and we have shown a construction of the neutral flavour baryon spectrum. All comparisons with observations agree at a percent or a sub percent level and should encourage further study within the model. In a further perspective we have suggested to look for a mesonic sector of the model based on a conjectured interaction term. More readily we suggest to look for exactly solvable constructions of protonic states and to discuss interpretations of the model.

\section*{Acknowledgments}
We thank Henrik Bohr and Mogens Stibius Jensen for interest in our Hamiltonian, Jane Hvolb\ae k Nielsen with colleagues at the Department of Physics for an inspiring institutional framework and Torben Amtrup for following the ups and downs through the years.

\end{document}